\documentclass[11pt]{article}
\usepackage{graphicx}
\usepackage{amsfonts}
\usepackage{amsmath}
\usepackage{amssymb}
\usepackage[margin=2.54cm]{geometry}

\begin{document}

\title{Practical approach to solvability: Geophysical application using complex decomposition into simple part (solvable) and \\ complex part (interpretable) for seismic imaging}

\author{August Lau and Chuan Yin \\
\\ Apache Corporation \\
        2000 Post Oak Blvd., Houston, Texas 77056 \\
        \\
        Email contact: \texttt{chuan.yin@apachecorp.com}}

\date{December 2, 2010}

\maketitle

\begin{abstract}

The classical approach to solvability of a mathematical problem is to define a method which includes certain rules of operation or algorithms.   Then using the defined method,  one can show that some problems are solvable or not solvable or undecidable depending on the particular method.  With numerical solutions implemented in a computer,  it might be more practical to define solvability of a mathematical problem as a complex decomposition problem.  The decomposition breaks the data into a simple part and a complex part.   The simple part is the solvable part by the method prescribed in the problem definition.  The complex part is the leftover of the simple part.   Complex part can be viewed as the "residual" of data or operator.  It should be interpreted and not to be discarded as useless.   We will give different examples to illustrate the more practical definition of solvability.   The complex part is not noise and should not be viewed as useless part of the data.  It has its own merit in terms of topological or geological interpretation.   We have de-emphasized absolute solvability and have emphasized the practical solvability where the simple and complex parts both play important roles.
\end{abstract}

\section*{Introduction}

We will start with some classical examples of solvability in mathematics.

The simplest example is the unsolvability of square root of 2 with rational numbers.   The prescribed method is to form fractions $m/n$ where $m$ and $n$ are integers and $n$ is nonzero.   Square root 2 is not solvable with this method of rational numbers or fractions.  This is normally proved by way of contradiction.   We can assume that square root 2 is a rational number.  Then find a contradiction in the end.   This is not a constructive proof.

The next example is solving polynomials.   From the Fundamental Theorem of Algebra,  it is known that a polynomial with complex coefficients could be factored into linear parts.   However,  the theorem is an existence theorem which is not constructive but only shows the existence of the linear factors.   

Before computers,  we had to solve the equation with only simple combinations.  We  restricted the definition of solvability of polynomials as solvable by radicals.   This means that we can only use radicals (square root,  cube roots,  squares, cubes etc) to form the solution.  For quadratic equation,  we have the quardratic formula which solves the quadratic equation with square roots and powers of the coefficients.  The quadratic is solvable by radicals. But the higher order polynomials are not solvable by radicals in general.  This leads to group theory,  more specifically Galois group theory of permutations,  which gives a more abstract formalism of solvable by radicals.

We then extend Galois group to more general group theory to explain physical problems as a group of transformations on physical equations.  The group theoretical application can be viewed as a group of symmetries or a group of operators like orthogonal matrices.  Then physics or other physical phenomenon is studied through the eyes of transformation groups.  It has been a successful approach to many physical problems.

The challenge to the group theoretic approach is that systems have to be described as reversible since each element of a group has an inverse.   We know that not all matrices are invertible.   For physical problems which are "lossy",   we have to introduce a more general formalism like semigroup.   The semigroup of transformations allows noninvertible elements or operators.   The collection of square matrices (e.g. all $2 \times 2$ matrices) form a semigroup under matrix multiplication but is not a group.  Semigroup can be used to view the world as non-reversible processes. 

\section*{Unsolvability of Cantor layers in seismic imaging}

Before we look into different ways to express simplicity/complexity, let us examine the Cantor layer construction in seismic imaging.   It is taking the Cantor middle third construction of layering.  See Lau and Yin SEG 2005 for more details.

\begin{figure}
\centering
  \includegraphics[width=6.6in]{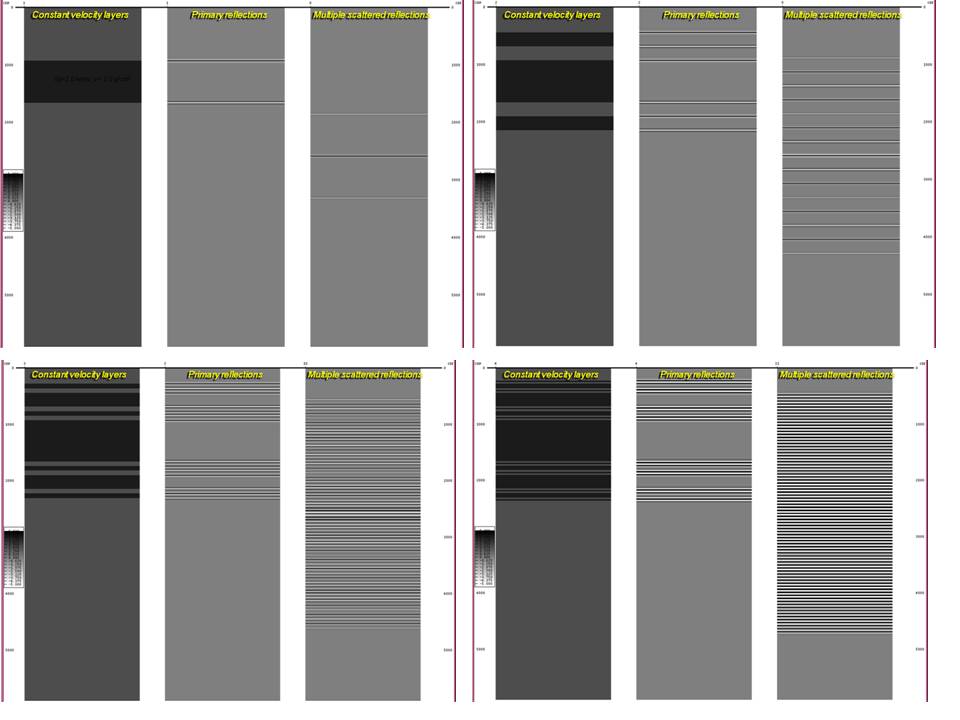}
\caption{different stages of Cantor layers and the corresponding primary and multiple reflections.}   
\label{fig1}
\end{figure}

Figure 1 shows different stages of Cantor layers and the corresponding primary and multiple reflections.   The Cantor layers converge to the Cantor set of reflections C.   The multiple reflections converge to a continuum, i.e.,

\begin{equation*}
C + C = I,
\end{equation*}
where $C$ is the Cantor set and $I$ is a continuum.

Cantor set is both sparse and dense.  If the Cantor set is the support of the primary reflection coefficients,  then $C + C$ is the support of the multiple scattered reflections which is a continuum.  Hence it is not possible to solve for Cantor coefficients which is sparse.  The multiples form a continuum which is the opposite of sparsity.

This is an example in which there is no solution.  So practical approach would be needed to simplify the problem.   

\section*{Reduce data complexity (with predetermined basis)}

The simplest way to reduce data is to do a linear filter based on a predetermined basis like Fourier sinusoids or  wavelet transform.  The complex decomposition is to apply the bandpass filter to output the simple part.   The difference of the input and output is the complex part.  The predetermined basis has the advantage of being fast computationally.  It has the disadvantage of imposing the geometry of the basis to the data.   So if Fourier transform is used,  the simple part would look like sinusoids.  Geologically it is not likely to have sinusoids which imply recurrent geologic events over many millions years.

\begin{figure}
\centering
  \includegraphics[width=6.5in]{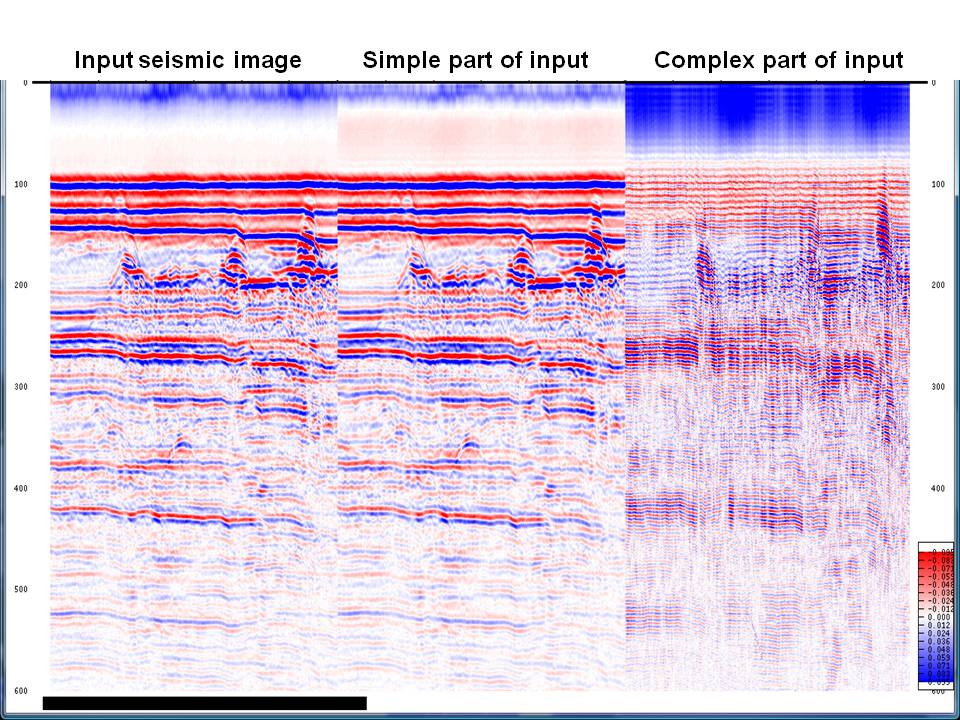}
\caption{The first panel is the input seismic image,  the second is the output of the simple part of the seismic and the last panel is the difference which is the complex part of the data.}
\label{fig2}
\end{figure}

Figure 2 shows 3 panels where the first panel is the input seismic image,  the second is the output of the simple part of the seismic and the last panel is the difference which is the complex part of the data.

\section*{Reduce data complexity (without predetermined basis)}

The smooth part of the data can be thought of as the simple part of the data.  We have used variational method to estimate the simple part of the data which can be used for further processing of seismic data.   We have employed the formulation of Chan et al for minimizing the $L1$ of the first derivative.  

Another approach is to use diffusion semigroup, e.g.,  Lau, Yin, Coifman and Vasilliou  SEG 2009.  We will show just the variational method in this paper.

\begin{figure}
\centering
  \includegraphics[width=6.5in]{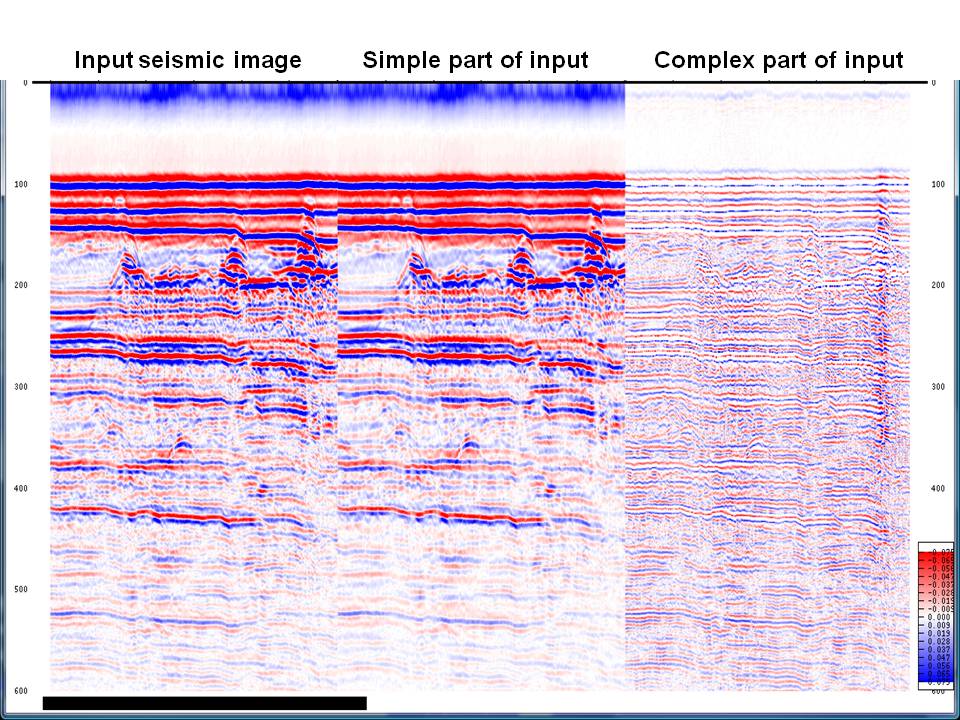}
\caption{The first panel is the input seismic image,  the second is the output of the simple part of the seismic and the last panel is the difference which represents the complex part of the data.}
\label{fig3}
\end{figure}

Figure 3 shows 3 panels where the first panel is the input seismic image,  the second is the output of the simple part of the seismic and the last panel is the difference which represents the complex part of the data.

Even though the third panel of the complex part looks noisy,  it is interpretable in terms of the details of the diffractions.   The geometry of the diffractions can be interpreted and it could be traced through the diffraction limps.  The actual amplitudes of the diffractions cannot be explained completely by a  wave equation.   This shows the geometric complexity of the diffractions.

The seismic data is not solvable in an absolute sense.  Not every sample can be explained by the wave equation.   The simple part is what is solvable with the current algorithm.  The complex part should not be discarded but should be interpreted.

\section*{Reduce operator complexity (with predetermined basis)}

Let us start by introducing a simple semigroup which is just the exponential.   We know that $\exp (t + s) = \exp (t) exp (s)$.    We can reduce wave equation from a second order partial differential equation into an exponential by performing Fourier transform on space and time domain.   Then the depth information is obtained by an exponential phase shift in Fourier domain and then transform to depth domain.   This is correct if it is simple velocity field like velocity field with no lateral variation.  This is commonly called fk migration in geophysics which is using Fourier ($f$ for frequency and $k$ for wavenumber) to solve the pde with predetermined basis like Fourier basis. Vertical variation is solved by recursive application of fk migration.

\begin{figure}
\centering
  \includegraphics[width=6.5in]{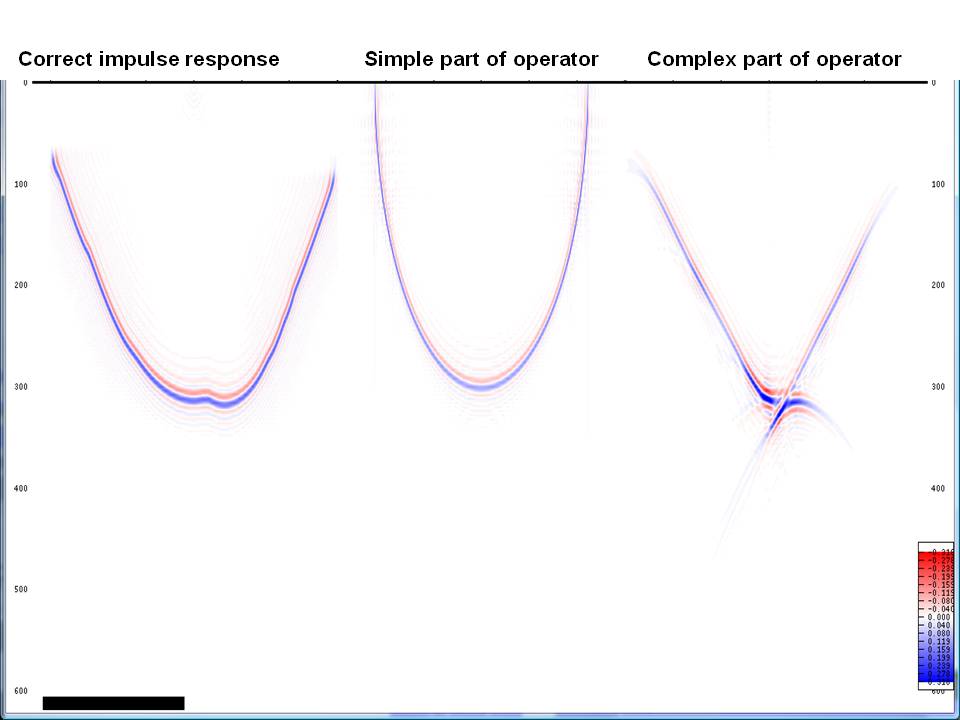}
\caption{Seismic responses to migrate the recorded data to their proper position.  The first panel is the correct operator generating the impulse response.  The second panel is the simple operator using a single velocity using fk migration.   The third panel is the complex part of the operator which is the "difference" of the first and second panel.}
\label{fig4}
\end{figure}

Figure 4 shows 3 panels of seismic responses to migrate the recorded data to their proper position.  The first panel is the correct operator generating the impulse response.  The second panel is the simple operator using a single velocity using fk migration.   The third panel is the complex part of the operator which is the "difference" of the first and second panel.

\section*{Reduce operator complexity (without predetermined basis)}

If we do not use a predetermined basis like Fourier transform,  we can use the velocity model coupled with the wave equation to form a simple operator.   See Sandberg and Beylkin 2009.   A less ambitious approach is to lump two variables into one,  e.g.,  velocity and thickness can be lumped into single variable of travel time.   See Lau and Yin SEG 2003 on "micro-tomography",  Lau and Yin arXiv 2010 on "Transformation Semigroup and Complex Topology".  Lumping variables is a way to reduce operator complexity without using a predetermined basis.

\section*{Reduce topological complexity}

One final approach is to reduce topological complexity.  We use computational topology to define Betti numbers.  The different Betti numbers of different dimensions indicate the number of "holes" for that dimension.  See Lau and Yin SEG 2009 on "topological simplicity" and arXiv 2010 on "L0+L1+L2 optimization".   Minimizing the Betti numbers is a way to reduce topological complexity.  By the same token,   topologically complex data cannot be solved by wave equation in an absolute sense.

\section*{Conclusion} 

We have used a more practical definition of solvability.  Absolute solvability is in general difficult to decide.  A data set or an operator could be reduced in complexity by various means as demonstrated above.  Complex decomposition breaks up the dataset or an operator into simple part (solvable part) and complex part (interpretable part).  The complex part should be preserved for interpretation and should not be discarded as "noise".   They have topological and geological meaning in terms of interpretation.  

\bibliographystyle{plain}
\bibliography{solvability}   

\begin{thebibliography}{1}

\bibitem{csv}
Chan, T., J. Shen, and L. Vese, 2003, Variational PDE Models in Image Processing: Notices of the American Mathematical Society, 50, Number 1, 14-26.

\bibitem{cllmnwz}
Coifman, R.R., S. Lafon, A. B. Lee, M. Maggioni, B. Nadler, F. Warner, S. W. Zucker, 2005, Geometric diffusions as a tool for harmonic analysis and structure definition of data:  Diffusion maps,  Proceedings of the National Academy of Sciences of the USA,  p. 7426-7431.

\bibitem{g}
Ghrist, R., Barcodes: The persistent topology of data, Bulletin of the American Mathematical Society, Vol. 45, Number 1, 2008, p. 61-75

\bibitem{l79}
Lau, A., 1979, Images of compact 0-dimensional semigroups, Colloquium Mathematicum, No. 2, Vol XL, p.219-222.

\bibitem{l80}
Lau, A., 1980,  Plane continua and transformation groups, Proceedings of the American Mathematical Society,  Vol. 78, No. 4, p. 608-610

\bibitem{ly2003}
Lau, A. and C. Yin, 2003,  Seismic prediction and micro-tomography: a case study of small traveltime error and waveform distortion, Internat. Mtg of Society of Exploration Geophysicists,  Expanded Abstracts,  p. 1774-1776.

\bibitem{ly2005}
Lau, A. and C. Yin, 2005,  Complex topology: its impact on seismic inversion and modeling in single component recording and multicomponent recording, Internat. Mtg of Society of Exploration Geophysicists,  Expanded Abstracts 1666-1668, Houston, Texas.

\bibitem{lrlpsg}
Lau, A.,  M. Roque-Sol, C. Lapilli, J. Perdomo, C. Shih, A. Gonzalez,  2008,  Imaging with complex decomposition: Numerical applications to seismic processing in difficult areas,  SEG Annual meeting expanded abstracts, p. 1986-1990.

\bibitem{ly09}
Lau, A. and C. Yin, 2009, Geometric simplicity as a migration criterion:  an application of computational topology to seismic imaging,  SEG Annual meeting expanded abstracts, p. 2773-2777.

\bibitem{ly10}
Lau, A., C. Yin , R.Coifman, A.Vassiliou  2009, Diffusion semigroups: A diffusion-map approach to nonlinear decomposition of seismic data without predetermined basis,  SEG Annual meeting expanded abstracts, p. 2327-2331.

\bibitem{ly10b}
Lau, A. and C. Yin,   2010,   L0+L1+L2 mixed optimization: a geometric approach to seismic imaging and inversion using concepts in topology and semigroup, arXiv:1007.1880v1.

\bibitem{ly10c}
Lau, A. and C. Yin,   2010,   Transformation Semigroup and Complex Topology: a study of inversion with increasing complexity, arXiv:1008.2668v1

\bibitem{sb}
Sandberg, K. and G. Beylkin  2009, Full-wave-equation depth extrapolation for migration, Geophysics, vol. 74, No. 6, November-December 2009, p. 121-128.

\end{thebibliography}

\end{document}